\documentstyle[12pt,axodraw,epsf,amssymb]{article}
\begin{document}
\baselineskip 0.65cm
\newcommand{\gsim}{ \mathop{}_{\textstyle \sim}^{\textstyle >} }
\newcommand{\lsim}{ \mathop{}_{\textstyle \sim}^{\textstyle <} }
\newcommand{\vev}[1]{ \left\langle {#1} \right\rangle }
\newcommand{\bra}[1]{ \langle {#1} | }
\newcommand{\ket}[1]{ | {#1} \rangle }
\newcommand{\EV}{ {\rm eV} }
\newcommand{\KEV}{ {\rm keV} }
\newcommand{\MEV}{ {\rm MeV} }
\newcommand{\GEV}{ {\rm GeV} }
\newcommand{\TEV}{ {\rm TeV} }
\def\diag{\mathop{\rm diag}\nolimits}
\def\Spin{\mathop{\rm Spin}}
\def\SO{\mathop{\rm SO}}
\def\O{\mathop{\rm O}}
\def\SU{\mathop{\rm SU}}
\def\U{\mathop{\rm U}}
\def\Sp{\mathop{\rm Sp}}
\def\SL{\mathop{\rm SL}}

%%%%%%%%%%
%%%%%%%%%%      title page
%%%%%%%%%%

\begin{titlepage}

\begin{flushright}
UT-962
\end{flushright}

\vskip 2cm
\begin{center}
{\large \bf  Supersymmetric Grand Unification Model with 
the Orbifold Symmetry Breaking in the Six Dimensional Supergravity}

\vskip 1.2cm
T.~Watari$^a$ and T.~Yanagida$^{a,b}$

\vskip 0.4cm
$^{a}$ {\it Department of Physics, University of Tokyo, \\
         Tokyo 113-0033, Japan}\\
$^{b}$ {\it Research Center for the Early Universe, University of Tokyo,\\
         Tokyo 113-0033, Japan}

\vskip 1.5cm
\abstract{We construct supersymmetric (SUSY) grand unification (GUT) 
models in the six dimensional space-time where the GUT symmetry 
is broken down to the standard-model gauge group 
by a simple orbifolding ${\bf T^2/Z_4}$ or ${\bf T^2/Z_6}$ 
and a pair of massless Higgs doublets in the SUSY standard model 
are naturally obtained. Since the background geometry here 
is simple compared with models using the Scherk-Schwarz mechanism, 
one might hope for an approximate gauge coupling unification 
in the present models. 
% Thus we consider that they are the first GUT models with the orbifold
%  breaking which may serve as a useful framework to 
%  investigate GUT dynamics in a higher dimensional space-time.
Here, the presence of the massless Higgs multiplets in the bulk 
is quite natural, since the anomaly cancellation in the six dimensional 
space-time requires N=2 hyper multiplets in the bulk,
some of which are origins of the Higgs doublets. 
However, the origin of the quarks and leptons is still not clear at all.}

\end{center}
\end{titlepage}

%%%%%%%%%%
%%%%%%%%%%      main part 
%%%%%%%%%%

% \section{Introduction}

Search for a solution to the doublet-triplet splitting problem in the
supersymmetric (SUSY) grand unification theory (GUT) has led us to
consider various extensions of the minimal SU(5)
SUSY-GUT\cite{mpm,DW,ss}. Recently, Kawamura\cite{kawamura1} has pointed 
out an interesting solution to this problem utilizing an ${\bf S}^1/{\bf
Z}_2$ orbifold 
in a five dimensional space-time, which is deeply related to the early 
suggestion by Witten\cite{witten}\footnote{There has been proposed 
another kind of solution to the doublet-triplet splitting problem in a
higher dimensional space-time\cite{imamura}.}. Although this original model is 
non SUSY-GUT, it is easily extended to the SUSY-GUT if one assumes an 
${\bf S}^1/({\bf Z}_2\vev{\sigma_1}\times {\bf Z}'_2\vev{\sigma_2})$ 
orbifold\cite{kawamura2,nomura1}. 
Here, this orbifold is also regarded as ${\bf R}^1/({\bf
Z}\vev{\sigma_1\sigma_2}\times'{\bf Z}_2\vev{\sigma_1})$, where the 
${\bf Z}\vev{\sigma_1\sigma_2}$ gives the symmetry-breaking boundary
condition a la Scherk-Schwarz\cite{sherk}.
A number of interesting features in this approach have been 
discussed\cite{many}. 
However, it is claimed\cite{march} recently that the Scherk-Schwarz 
breaking is equivalent to the Wilson line breaking\cite{hosotani}
localized at a fixed point. 
Thus, % although the Wilson line SU(5) breaking is absent in the bulk,
the nontrivial background of gauge field exists at the fixed point, 
and its effect to the tree level gauge coupling is incalculable
and it may be beyond the
naive dimensional analysis\cite{nomura1}\footnote{In the string theory 
the SU(5) gauge symmetry is realized on the branes where five D-branes 
coincide, and the Wilson line breaking corresponds to a spatially
parallel separation 
of the five D-branes in a T-dual manifold\cite{Tdual}. 
The Wilson line breaking localized at fixed points implies 
that transverse fluctuation modes of D-branes
(including their winding modes) develop expectation values
differently on the three and two D-branes. 
Therefore, there is no reason that the expectation value of dilaton 
takes the same value at the ``positions'' of the separated 
SU(3) three branes and SU(2) two branes. 
Since the values of dilaton correspond to the gauge couplings, 
the coupling constants of each gauge groups are, in general, 
different from each others.}.   
Therefore, it is not necessarily obvious to maintain the gauge 
coupling unification even at the tree level.
 
The purpose of this letter is to show that a SUSY-GUT model
with the orbifold GUT breaking mechanism 
is constructed in the six dimensional space-time without the
Scherk-Schwarz breaking\footnote{There has been,
recently, discussed the SO(10) GUT in the six dimensional
space-time\cite{asaka}. However, the Scherk-Schwarz boundary breaking 
is still postulated.}. In this letter, we use the term 
``orbifold GUT breaking model'' as a class of models that do not use 
the Scherk-Schwarz (or boundary breaking) mechanism.
% Then, in the absence of the Wilson line background that explicitly 
% distinguishes the color SU(3) and the flavor SU(2) subgroups,
% the common gauge coupling constant of the gauge multiplets  
% may be justified at the tree level due to the residual SU(5) gauge 
% symmetry noticed by Hall and Nomura\cite{nomura1}. 
Since the background geometry here is simple compared with models using
the Scherk-Schwarz mechanism, one might hope for an approximate gauge
coupling unification in the present models.
We consider a simple orbifold ${\bf T}^2/{\bf Z}_4$. 
We find that only a pair of the Higgs doublets $H_{\rm f}$ and 
${\bar H}_{\rm f}$ survive together with the gauge multiplets of
SU(3)$_{C}\times$SU(2)$_{L}\times$U(1)$_{Y}$ in the  
extra two dimensional bulk. 
The matter multiplets ${\bf 5^*}$ and ${\bf 10}$ are assumed to reside on 
a fixed point that preserves SU(5) symmetry. 
We also find that a similar model can be constructed 
on another simple orbifold ${\bf T}^2/{\bf Z}_6$.

We first note that the five dimensional space-time is too small to have the 
desired orbifold GUT breaking model (without the  Scherk-Schwarz
mechanism). 
% There are two reasons. The first one is that 
% the two fixed points of the ${\bf S}^1/{\bf Z}_2$ orbifold 
% has exactly the same property, and there is no room for the fixed point
% that preserve the (four dimensional) SU(5) symmetry. 
% The other is that 
The orbifold projection associated with the 
${\bf S}^1/{\bf Z}_2$ compactification is not enough to eliminate 
all the unwanted particles contained in the N=2 SUSY SU(5) 
multiplets. Such an elimination is possible only by using the
Scherk-Schwarz mechanism in the ${\bf S}^1/({\bf Z}_2 \times {\bf Z}'_2)
\simeq {\bf R}^1/({\bf Z}\times' {\bf Z}_2)$ compactification.
Therefore, we consider the orbifold GUT breaking model 
in the six dimensional space-time. Furthermore, there is another reason
to assume the higher dimensional space-time; the R-symmetry that is a
crucial and unique symmetry to forbid a  constant term in the
superpotential\cite{izawa-yanagida} may arise from a rotation
symmetry in the extra space. 
In this sense the present analysis provides the first basic and natural 
GUT model in a higher dimensional space-time\footnote{The present 
models will serve as a useful framework for various investigations
(including the confirmation of the gauge coupling unification) 
of orbifold GUT breaking models in a higher dimensional space-time.}. 
We restrict ourselves to the six dimensional supergravity in this letter,
since if we go further the consistency condition becomes more
restrictive. We do not consider that the effective field theory should be
necessarily formulated in the ten dimensional supergravity, even if 
the underlying fundamental theory is given by the string theory.

The possible rotational symmetry that 
the two dimensional torus ${\bf T}^2$ in the extra dimensional space 
possesses is ${\bf Z}_2$, ${\bf Z}_3$, ${\bf Z}_4$ or ${\bf
Z}_6$. These are the symmetries that rotate the 4th-5th plane by $\pi$,
$2\pi/3$, $\pi/2$ and $\pi/3$, respectively.
We require that the orbifold projection condition explicitly
distinguishes the color SU(3) and the flavor SU(2) of the SU(5)  
indices and breaks the SU(5) GUT down to the standard-model gauge group. 

The N=2 SU(5) vector multiplet propagates in the six dimensional bulk. 
The lengths $L$ of the extra dimensions are assumed to be of order of 
% the inverse of the GUT scale 
1/(GUT scale) and 
the fundamental scale of the theory $M_*$ is of order $10^{17}\GEV$.
This lower cut-off scale is a crucial point in keeping the
approximate gauge coupling unification at the effective GUT scale 
$\sim 10^{16}\GEV$\cite{nomura1,rattazzi}. 
The volume of the extra two dimensional space\footnote{This 
large extra dimensions may be welcome to the gaugino mediation 
scenario of SUSY breaking\cite{kaplan}.} is about $(M_*L)^2 \sim
10^{2}$.
% \cite{nomura1}

Now that the SU(5) vector multiplet propagates in the six dimensional bulk,
the six dimensional box anomaly must be canceled.
Six dimensional box anomalies consist of pure gauge anomalies tr($F^4$)
and (tr($F^2$))$^2$, a gauge-gravity mixed anomaly tr($F^2$)tr($R^2$) and
pure gravitational anomalies\cite{anomaly6D}.
Among these, pure gravitational anomalies can be canceled by
introduction of SU(5) singlet fields and reducible anomalies
(tr($F^2$))$^2$ and tr($F^2$)tr($R^2$) can be canceled by the
Green-Schwarz mechanism\cite{GS6}. Thus we only care about 
the pure gauge tr($F^4$) anomaly of the SU(5) gauge theory.
In (1,0)-SUSY (N=2 SUSY in four dimensional sense) gauge theories 
in six dimensions, N=2 hyper multiplets have chirality opposite to
that of the N=2 vector multiplet.
Therefore, the pure gauge box anomaly from the gauge fermions of the 
N=2 SU(5) vector multiplet can be canceled by introducing 
N=2 hyper multiplets. 
The anomaly of the SU($n$) N=2 vector multiplet is  $-2n$ times that of
the N=2 hyper multiplet in SU($n$)-fundamental
representation (up to reducible anomaly)\cite{anomaly6D}, and hence 
we introduce ten (${\bf 5}$+${\bf 5}^*$) hyper multiplets in the six
dimensional bulk\footnote{The anomaly from the N=2 vector multiplet is
completely canceled by the N=2 hyper multiplet in the adjoint
representation. Here, they may form a vector multiplet of the N=4 SUSY.
Additional three (${\bf 5}+{\bf 5}^*$) hyper multiplets and a (${\bf
10}+{\bf 10}^*$) hyper multiplet do not give rise to an irreducible pure 
gauge anomaly, and these multiplets may be of phenomenological use. In
this letter, however, we only discuss the simplest possibility 
(ten ${\bf 5}+{\bf 5}^*$ in the text).}. 
% 10 + 10^* representation has anomaly ... and possibility ....

We expect that the four dimensional N=1 vector multiplets of the 
SU(3)$_{C}\times$SU(2)$_{L}\times$ U(1)$_{Y}$ in the minimal
SUSY standard model (MSSM) comes from the above N=2 SU(5)
vector multiplet.
Since we introduced N=2 hyper multiplets in ${\bf 5}$+${\bf 5}^*$ 
representation, it is possible that the $H_{\rm f}$ and $\bar{H}_{\rm
f}$ N=1 chiral multiplets in the MSSM also originate from the bulk fields.  
Spectrum of massless particles that live in the bulk  are determined
by the orbifold projection condition that distinguishes the color SU(3) 
and the flavor SU(2), and hence the remaining particles may be chosen as 
only doublets by adopting an appropriate orbifold projection as shown below. 
The phenomenological reason for which we identify the Higgs as bulk
fields (not as fixed point fields), will be explained later.
On the contrary, we postulate by hand that the quarks and leptons 
N=1 chiral multiplets, (${\bf 5}^*+{\bf 10}$), reside on orbifold 
fixed points. 
This is the weakest point in the present approach.

In the model construction of the orbifold GUT breaking, the  
existence of the fixed point that preserves the SU(5) symmetry is 
an important ingredient\cite{nomura1}.
A natural explanation of the anomaly cancellation in terms of the SU(5)
GUT which is nothing but a miracle in the standard model,
the charge quantization of the U(1)$_{Y}$, and the bottom-tau Yukawa
unification are the major reasons that we believe SUSY-GUT 
along with the gauge coupling unification suggested from the
experiments. The above three features are still maintained 
if the orbifold geometry has a fixed point 
that preserves the four dimensional SU(5) GUT symmetry
even though the orbifold GUT breaking model has no complete higher
dimensional SU(5) symmetry.
We consider the model in which the three families of quarks and leptons 
${\bf 5}^*$ + ${\bf 10}$ reside on such a fixed point. 

In order to have a fixed point which preserves the SU(5) symmetry, 
the orbifold group must have nontrivial and proper subgroup.
First, the isotropy group associated to such a fixed point\footnote{
Isotropy (sub)group of a point is a subgroup of a transformation group,
say the orbifold group, which consists of elements that fix the point.} 
is not trivial by definition. 
Secondly, if the isotropy group were identical to the whole orbifold
group, then the orbifold projection associated to the isotropy group 
of such a % an SU(5) preserving 
fixed point would distinguish the color SU(3) and the flavor SU(2)
subgroups of the SU(5), and hence the SU(5) symmetry would not be preserved.
Therefore, the isotropy group of such a fixed point is nontrivial and
proper subgroup of the whole orbifold group. 
This means that the orbifold group candidate are ${\bf Z_4}$ and ${\bf
Z_6}$ since ${\bf Z}_2$ and ${\bf Z}_3$ do not have such a subgroup. 
Therefore, we consider the ${\bf T}^2/{\bf Z}_4\vev{\sigma}$ and ${\bf T}^2/
{\bf Z}_6\vev{\sigma}$ model where the $\sigma$ is the generator of the
each orbifold group ${\bf Z}_4$ and ${\bf Z}_6$.

Let us first consider the model on ${\bf T}^2/{\bf Z}_4\vev{\sigma}$
orbifold.
The generator $\sigma$ of the ${\bf Z}_4$ transformation rotates
the 4th-5th plane by $\pi/2$ :
\begin{equation}
 (z\equiv (x_4+ix_5))\rightarrow \sigma \cdot z 
                            = e^{i(\theta=\frac{2\pi}{4})} z.
\end{equation}
The geometric picture of the ${\bf T}^2/{\bf Z}_4\vev{\sigma}$ 
orbifold is shown in Fig.\ref{fig:Z4}. 
There are two ${\bf Z}_4\vev{\sigma}$ fixed points and 
there is one ${\bf Z}_2\vev{\sigma^2}$ fixed point\footnote{In this
letter, we call the fixed point whose isotropy group is ${\bf
Z}_2\vev{\sigma^2}$ as ${\bf Z}_2\vev{\sigma^2}$ fixed point for
brevity. Similar terminology is also used later in the ${\bf T}^2/{\bf Z}_6$
orbifold model.}.
We can identify the latter fixed point as the SU(5) preserving fixed
point, as seen below.
The N=2 bulk fields consist of an N=1 SU(5) vector multiplet $V(z)$, 
an N=1 chiral multiplet $\Sigma(z)$ that transforms as an adjoint 
representation of the SU(5), and N=1 chiral multiplets 
$F_j(z)$ and $\bar{F}^j(z)$ (j=1,...,10) that transforms 
${\bf 5}$ and ${\bf 5}^*$ before the orbifold projection.
The massless spectrum of the gauge theory on the orbifold is given 
by the zero modes that
satisfy the following orbifold projection condition:
\begin{equation}
V(x,z) = \widetilde{\gamma}_{\sigma} V(x,e^{i\theta}z) 
         \widetilde{\gamma}_{\sigma}^{-1}, \quad
\Sigma(x,z) = e^{i\theta} \widetilde{\gamma}_{\sigma} 
             \Sigma(x,e^{i\theta}z) \widetilde{\gamma}_{\sigma}^{-1},
\label{eq:OP4-1}
\end{equation}
\begin{equation}
F(x,z)_j = e^{i\theta n_j} \widetilde{\gamma}_{\sigma} F(x,e^{i\theta}z)_j ,
\quad
\bar{F}(x,z)^j = e^{i\theta(-1-n_j)} 
            \widetilde{\gamma}_{\sigma}^{-1} \bar{F}(x,e^{i\theta}z)^j,
\label{eq:OP4-2} 
\end{equation}
where $\theta = (2 \pi)/4$. 
Rotational charges $n_j$ for the hyper multiplet can be 
$n/2 ~(n=0,1,2,3,...)$. 
$\widetilde{\gamma}_{\sigma}$ is the (5 $\times$ 5) 
gauge twisting matrix associated to the generator $\sigma$ 
that must satisfy $(\widetilde{\gamma}_{\sigma})^4 ={\bf 1}$.

We take the gauge twisting matrix $\widetilde{\gamma}_{\sigma}$ as
\begin{equation}
\widetilde{\gamma}_{\sigma}= 
\diag(e^{i \theta m},e^{i \theta m},e^{i \theta m},
      e^{i (\theta m +\pi)},e^{i (\theta m +\pi)}),
\label{eq:twist4}
\end{equation}
where $m$ is an arbitrary integer and $\theta = (2 \pi)/4$. Under this
choice, the ${\bf Z}_2\vev{\sigma^2}$ fixed point preserves the SU(5)
symmetry because the gauge twisting matrix
$\widetilde{\gamma}_{\sigma^2} \equiv 
(\widetilde{\gamma}_{\sigma})^2 \propto {\bf 1}$ does not make any
discrimination between the color SU(3) and the flavor SU(2).
We put the three families of quarks and leptons on this fixed point.
 
Now we can see that the massless particles (Kaluza-Klein zero modes) 
from the N=2 SU(5) vector multiplet are just the N=1 vector multiplets 
of the MSSM.
Only one pair of the N=1 chiral multiplets $H_{\rm f}$ from 
the $F_1$ and the $\bar{H}_{\rm f}$ from the $\bar{F}^2$ survive 
the orbifold projection conditions Eq.(\ref{eq:OP4-2})
if we take $n_1 =(2-m)$, $n_2 =(1-m)$ and $n_j$ (j = 3,...,10) to be half
integers. 
These are exactly the pair of Higgs doublets in the MSSM. 
No other unwanted particle remains massless.
We can also see that the triangle anomalies which might
appear at fixed points because of the orbifolding completely 
vanish\cite{ACG}. 

We obtain the desired massless Higgs multiplets  $H_{\rm f}$ 
and $\bar{H}_{\rm f}$ in the bulk. As a matter of fact, this is a 
necessary property as long as
the third family of the quarks and leptons reside on the SU(5)
preserving fixed point. The reason is the following.
Suppose that the two Higgs doublets reside on one of the two 
${\bf Z}_4\vev{\sigma}$ fixed points and the third family reside on the
${\bf Z}_2\vev{\sigma^2}$ fixed point. Let us consider how the
Yukawa couplings in the superpotential are generated.
Since the quark and lepton multiplets are separated from the two Higgs
doublets by the distance $M_*L \sim 10$, 
an exchange of particles of mass of order
of the fundamental scale $M_*$ is not enough to induce the Yukawa
couplings because of the damping of the wave function $e^{-M*L}
\lsim 10^{-4}$.
Only the Kaluza-Klein particles of the ${\bf 5}$+${\bf 5}^*$ hyper
multiplets can do the job. However, we can see that all those 
Kaluza-Klein particles have zero wave function at the ${\bf Z}_4$ 
fixed points in models where no massless Higgs multiplet remains
in the bulk. Therefore, necessary Yukawa couplings are not generated 
by the exchanges of the Kaluza-Klein particles. 

It is easy to see that a similar argument to the  above 
also holds in the model of ${\bf T}^2/{\bf Z}_6\vev{\sigma}$ orbifold.
The generator $\sigma$ rotates the 4th-5th plane by $\pi/3$:
\begin{equation}
 z \rightarrow e^{i \theta} z \qquad \qquad 
                        \left(\theta= \frac{2\pi}{6}\right).
\end{equation}
Orbifold projection conditions are the same as Eq.(\ref{eq:OP4-1}) and
Eq.(\ref{eq:OP4-2}) with $\theta=(2\pi)/4$ replaced by $\theta=(2 \pi)/6$. 
The (5 $\times$ 5) gauge twisting matrix associated to the generator
$\sigma$ can be given by
\begin{equation}
\widetilde{\gamma}_{\sigma}= 
\diag(e^{i \theta m},e^{i \theta m},e^{i \theta m},
      e^{i (\theta m +\pi)},e^{i (\theta m +\pi)}),
\end{equation}
as in Eq.(\ref{eq:twist4}), or by
\begin{equation}
\diag(e^{i \theta m},e^{i \theta m},e^{i \theta m},
      e^{i (\theta m \pm 2\pi /3)},e^{i (\theta m \pm 2\pi/3)}),
\end{equation}
where $m$ is again an arbitrary integer.
In the former case ${\bf Z}_3\vev{\sigma^2}$ fixed point are the SU(5)
preserving fixed point and in the latter case the ${\bf
Z}_2\vev{\sigma^3}$ fixed point preserves the SU(5) (see Fig.\ref{fig:Z6}). 
Precisely the N=1 vector multiplets of the MSSM survive the orbifold
projection, and the two massless Higgs doublets remain in the bulk
if we take
the rotational charges $n_j$ of the N=2 hyper multiplets as 
$n_1 = (-m -3)$, $n_2 =(-m +2)$ and $n_j =(-m -2),(-m-5)$, (half integers)
for (j=3,...,10) in the former case, and 
$n_1=(-m \mp 2)$, $n_2=(-m \mp 2-1)$ and $n_j = (-m \mp 1 -4)$, $(-m \mp 1
-3)$, (half integers) for (j=3,...,10) in the latter case.
We also see that the two Higgs doublets should be in the bulk 
to have the sufficiently large Yukawa couplings 
in this ${\bf T}^2/{\bf Z}_6$ orbifold model.

% The argument given above that the two Higgs doublets in the bulk is a
% necessary property for the sufficient Yukawa couplings also (perhaps??) 
% holds in this ${\bf T}^2/{\bf Z}_6$ orbifold model.

If we assume that the second and/or the first family also reside on the
SU(5) preserving fixed point, then we have to find some mechanism to break
the SU(5) relation $m_s=m_{\mu}$ and/or $m_d=m_e$. This may be
realized through the mixing of these quarks and leptons
with the massive multiplets that propagate in the bulk. In the  
${\bf T}^2/{\bf Z}_6$ orbifold model such massive multiplets may be
supplied by Kaluza-Klein towers of the hyper multiplets $F_j$ and
${\bar F}^j$, and in the ${\bf T}^2/{\bf Z}_4$ orbifold model they are
contained in the heavy particles at the cut-off scale $M_*$. A detailed
phenomenological aspects of the present models will be given elsewhere. 

\section*{Acknowledgments}
We are grateful to Dr.Yasunori Nomura for useful discussion.
T.W. thanks the Japan Society for the Promotion of Science for
financial support.
This work was partially supported by ``Priority Area: Supersymmetry and
Unified Theory of Elementary Particles (\# 707)'' (T.Y.).

\begin{figure}[h]
\begin{picture}(200,200)(-200,-50)
\Line(-30,0)(130,0) \Line(-30,100)(130,100)
\Line(0,-30)(0,130) \Line(100,-30)(100,130)
\Vertex(0,0){3} \Text(4,-4)[lt]{${\bf Z}_4$} 
\Vertex(50,50){3} \Text(54,46)[lt]{${\bf Z}_4$}
\CArc(0,50)(3,0,360) \Text(-4,46)[rt]{${\bf Z}_2$}
\CArc(50,0)(3,0,360) \Text(54,-4)[lt]{${\bf Z}_2$}
\ArrowLine(25,25)(0,50) \ArrowLine(25,25)(50,0)
\end{picture}
\caption{${\bf T}^2/{\bf Z}_4$ orbifold geometry is described. There are 
 two ${\bf Z}_4\vev{\sigma}$ fixed points ($\bullet$'s in the figure) and one
 fixed point ($\circ$ in the figure) whose isotropy group is 
${\bf Z}_2\vev{\sigma^2}$. The arrow denotes the identification between
 mirror images under ${\bf Z}_4/{\bf Z}_2$. This ${\bf Z}_2\vev{\sigma^2}$ 
fixed point is the SU(5) preserving fixed point.}
\label{fig:Z4}
\begin{picture}(200,200)(-170,-50)
\Line(-30,0)(130,0) \Line(20,87)(180,87)
\Line(-15,-26)(65,113) \Line(85,-26)(165,113)
\Vertex(0,0){3} % \Text(4,-4)[lt]{${\bf Z}_3$} 
\Vertex(50,29){3} \Text(54,25)[lt]{${\bf Z}_3$} 
\Vertex(100,58){3} \Text(104,54)[lt]{${\bf Z}_3$} 
\ArrowLine(75,43)(100,58) \ArrowLine(75,43)(50,29)
\CArc(0,0)(4,0,360) \Text(-4,4)[rb]{${\bf Z}_6$}
\CArc(50,0)(4,0,360) \Text(46,-4)[rt]{${\bf Z}_2$}
\CArc(25,43)(4,0,360) \Text(21,48)[rb]{${\bf Z}_2$}
\CArc(75,43)(4,0,360) \Text(71,48)[rb]{${\bf Z}_2$}
\ArrowLine(50,0)(25,43) \ArrowLine(25,43)(75,43) \ArrowLine(75,42)(50,0)
\end{picture}
\caption{Geometry of ${\bf T}^2/{\bf Z}_6$ is described. This orbifold
 has three fixed points whose isotropy groups are all different:
namely the $\bullet$(${\bf Z}_3\vev{\sigma^2}$ fixed), the 
$\circ$(${\bf Z}_2\vev{\sigma^3}$ fixed) and 
the $\bullet$-$\circ$(${\bf Z}_6\vev{\sigma}$ fixed) in this figure. 
Each of the fixed point $\bullet$ and $\circ$ can be an SU(5) preserving 
fixed point.
Arrows denote the identification between mirror images under 
${\bf Z}_6/{\bf Z}_3$ ($\bullet$'s) and ${\bf Z}_6/{\bf Z}_2$ ($\circ$'s), 
respectively.}
\label{fig:Z6}
\end{figure}
\end{document}